\documentclass[useAMS,usenatbib]{mn2e}
\usepackage{amssymb}
\usepackage{graphicx}

\def \curlB {\vec{\nabla}\times (e^\nu \vec{B})}

\title{CCOs and the hidden magnetic field scenario}
\author[D.~Vigan\`o and J.A.~Pons]{D.~Vigan\`o$^{1}$\thanks{E-mail: daniele.vigano@ua.es} and J.A.~Pons$^{1}$\\
$^{1}$Departament de F\'{\i}sica Aplicada, Universitat d'Alacant, Ap. Correus 99, 03080 Alacant, Spain}
\begin{document}

\date{}
\pagerange{\pageref{firstpage}--\pageref{lastpage}}
\maketitle

\label{firstpage}

\begin{abstract}
CCOs are X$-$ray sources lying close the center of supernova remnants, with inferred values of the surface magnetic fields significantly lower ($\lesssim 10^{11}$~G) than those of standard pulsars.
In this paper, we revise the hidden magnetic field scenario, presenting the first 2D simulations of the submergence and reemergence of the magnetic field in the crust of a neutron star. A post-supernova accretion stage of about $10^{-4}$-$10^{-3} M_\odot$ over a vast region of the surface is required to bury the magnetic field into the inner crust.
When accretion stops, the field reemerges on a typical timescale of 1-100 kyr, depending on the submergence conditions. After this stage, the surface magnetic field is restored close to its birth values. A possible observable consequence of the hidden magnetic field is the anisotropy of the surface temperature distribution, in agreement with observations of several of these sources. We conclude that the hidden magnetic field model is viable as alternative to the anti-magnetar scenario, and it could provide the missing link between CCOs and the other classes of isolated neutron stars.
\end{abstract}

\begin{keywords}
stars: neutron -- stars: magnetic field -- stars: pulsars: general
\end{keywords}

\section{Introduction}
The handful of reported Central Compact Objects (CCOs) forms a class of X-ray sources (see \cite{deluca08,halpern10} for recent reviews), located close to the center of $\sim$kyr old supernova remnants. CCOs are supposed to be young, isolated, radio-quiet neutron stars (NSs). They show very stable, thermal-like spectra, with hints of temperature anisotropies, in terms of large pulsed fraction, or small emitting regions for hot ($0.2$-$0.4$ keV) blackbody components.

The period is known for only three cases: $P=424$ ms for 1E 1207.4-5209 in G296.5+10.0 \citep{zavlin00}, $P=112$ ms for RX J0822.0-4300 in Puppis A \citep{gotthelf09}, and $P=105$ ms for CXOU J185238.6+004020 in Kes 79 \citep{gotthelf05} (hereafter 1E 1207, Puppis A and Kes 79, respectively). For Kes 79, \cite{halpern10} reported a period derivative of $\dot{P}\simeq 8.7\times 10^{-18}$ ss$^{-1}$; for 1E 1207, \cite{halpern11} give two equally good timing solutions, with $\dot{P}=2.23\times 10^{-17}$ ss$^{-1}$ and $\dot{P}=1.27\times 10^{-16}$ ss$^{-1}$. Only an upper limit $\dot{P}<3.5\times 10^{-16}$ ss$^{-1}$ is available for the CCO in Puppis A \citep{gotthelf10,deluca12}. Applying the classical dipole-braking formula gives an estimate for the dipolar component of the \textit{external} magnetic field (MF) of $B_p\sim 10^{10}-10^{11}$~G (at the pole). This low value has led to the interpretation of CCOs as ``anti-magnetars'', i.e., NSs born with very low MFs, which have not been amplified by dynamo action due to their slow rotation at birth \citep{bonanno05}.

An alternative scenario to the ``anti-magnetar'' model that explains the low values of $B_p$ is to consider the fallback of the debris of the supernova explosion onto the newborn NS. During a time interval of few months after the explosion, it accretes material from the reversed shock, at a rate far superior than Eddington limit \citep{colgate71,blondin86,chevalier89,houck91,colpi96}. This episode of hypercritical accretion could bury the MF into the NS crust, resulting in an external MF (responsible for the spindown of the star) much lower than the internal ``hidden" MF.

When accretion stops, the screening currents localized in the outer crust are dissipated on Ohmic timescales and the MF eventually reemerges. The process of reemergence has been explored in past pioneer works \citep{young95,muslimov95,geppert99} with simplified 1D models and always restricted to dipolar fields. It was found that, depending on the depth at which the MF is submerged (the submergence depth), it diffuses back to the surface on radically different timescales $10^3-10^{10}$ yr. Thus, the key issue is to understand the submergence process and how deep can one expect to push the field during the accretion stage. This latter issue has only been studied (also in 1D and for dipolar fields) by \cite{geppert99}, in the context of SN 1987A. They conclude that the submergence depth depends essentially on the total accreted mass. More recently, \cite{ho11} has revisited the same scenario in the context of CCOs, using a 1D cooling code and studying the reemergence phase of a buried, purely dipolar field,  with similar conclusions to previous works. The {\it hidden magnetic field scenario} has also been proposed by \cite{shabaltas12} for the CCO in Kes 79. They can explain the observed high pulsed fraction ($f_p\approx 60\%$) of the X$-$ray emission with a sub-surface MF of $\approx 10^{14}$~G, that causes the required temperature anisotropy.

In this paper, we further explore the viability of this scenario with improved calculations that can account for temperature anisotropies. We present the first results from 2D simulations of the submergence and rediffusion of the MF, using the recent extension \citep{vigano12} of the magneto-thermal evolution code of \cite{pons09}. The new code is able to follow the coupled long-term evolution of MF and temperature including the non-linear Hall term in the induction equation. This allows to follow the evolution for any MF geometry (not restricted to dipoles) and includes state-of-the-art microphysical inputs.

%%%%%%%%%%%%%%
\begin{figure}
 \centering
\includegraphics[width=.2\textwidth]{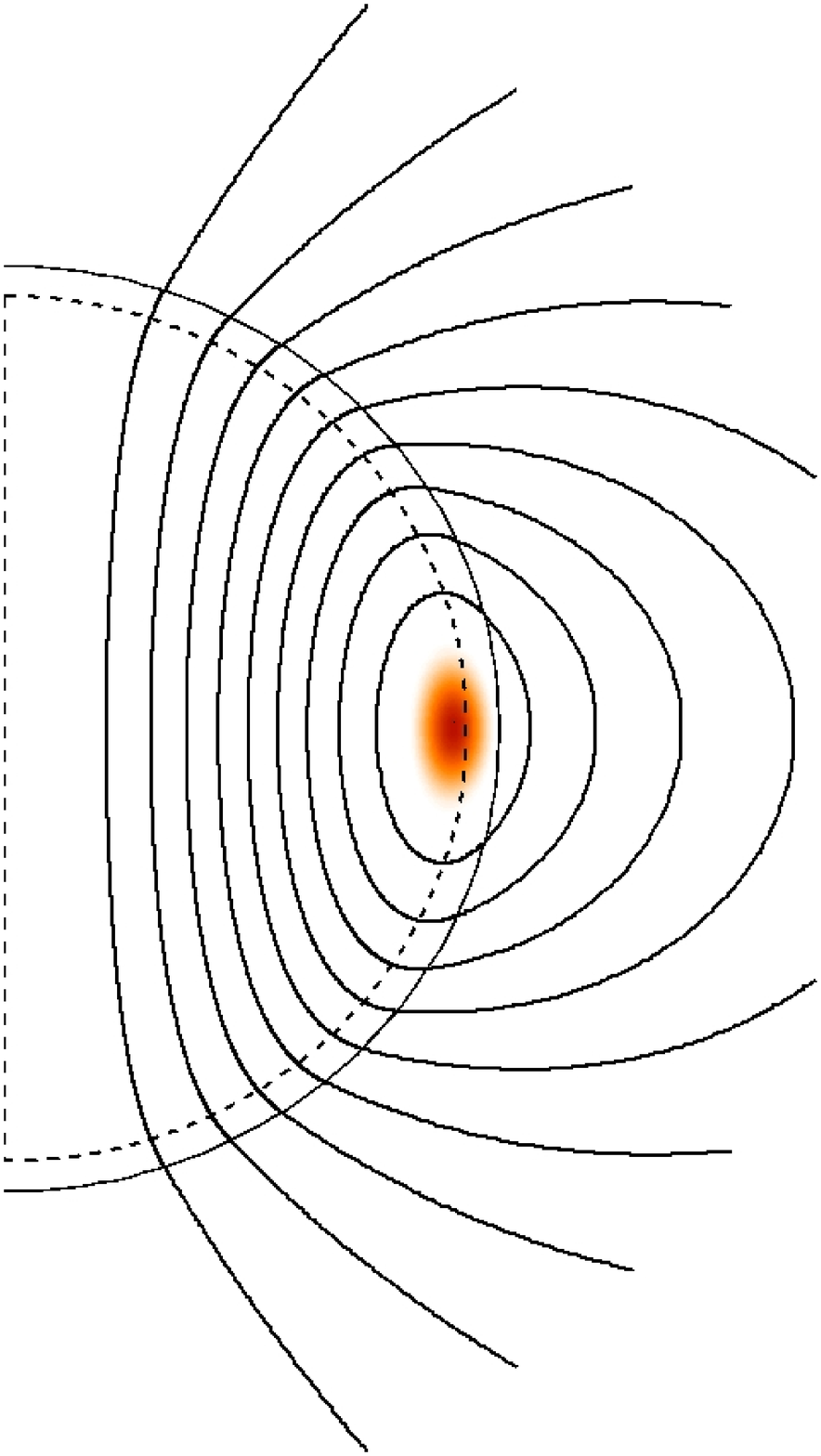}
\includegraphics[width=.2\textwidth]{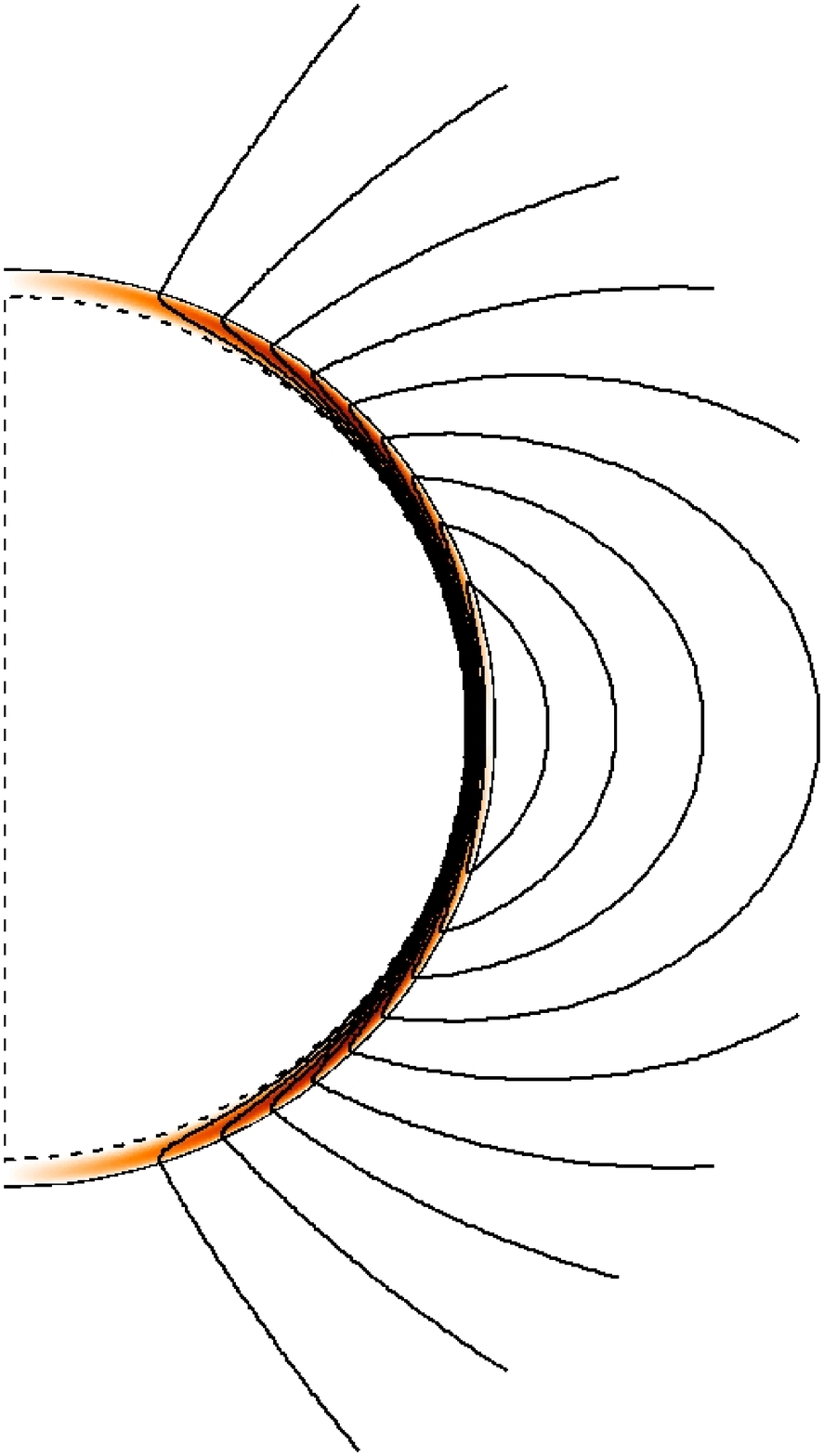}\\
\includegraphics[width=.22\textwidth]{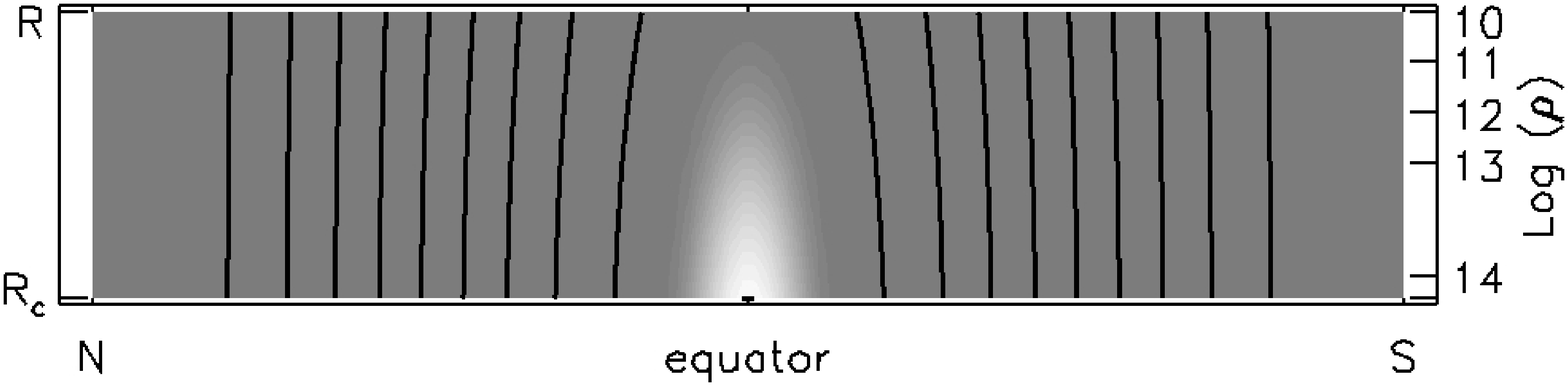}
\includegraphics[width=.22\textwidth]{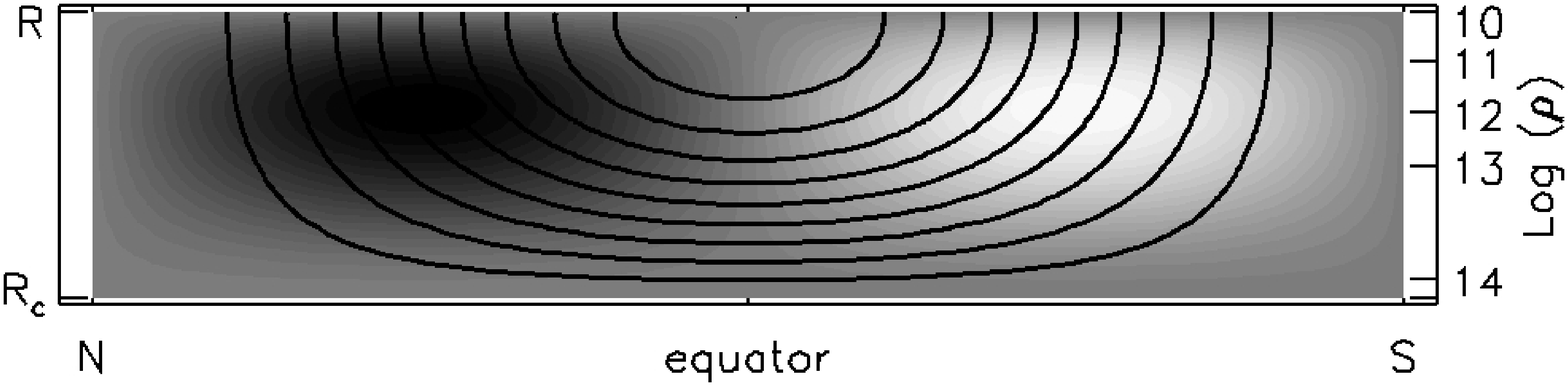}
\caption{Initial configuration of model A (left) and model B (right). Poloidal field lines (solid) smoothly match a vacuum solution outside the star. Upper panels: global configuration, the dashed line indicates the core-crust interface (denoted by $R_c$); the color scale represents the toroidal MF intensity. Lower panels: planar representation of the MF in the crust ($r\in[R_c,R]$), as a function of the polar angle $\theta$; the grey scale is normalized to the maximum value of $B_t$, and the white/black regions indicate positive/negative values.}
 \label{fig_initial}
\end{figure}

%%%%%%%%%%%%%%%%%%%%%%%%%%%%%
\section{Basic equations and initial model}

We assume spherical symmetry for our background star (very small deformations induced by the MF or rotation are neglected) with redshift and other metric factors only depending on the radial coordinate, and we adopt the standard static metric 
\begin{equation}
ds^2= -c^2 e^{2\nu(r)}dt^2+ e^{2\lambda(r)}dr^2 + r^2 d\Omega^2~,
\label{metric}
\end{equation}
where $ e^{-\lambda(r)} = (1 - 2 G m(r)/c^2 r)^{1/2}$. The density, pressure, enclosed mass $m(r)$, and $\nu(r)$ profiles are obtained by solving the hydrostatic equilibrium equations. We consider a 1.4 $M_\odot$ NS (see Section 2 in \cite{pons09}, Section 4 of \cite{aguilera08} and references within for all details about the equation of state and other microphysical input), with two different initial configuration, shown in Fig.~\ref{fig_initial}. Model A has an initial dipolar component of the MF extended to the core, with a toroidal component to mimic a typical MHD equilibrium configuration \citep{ciolfi09,lander09}, in which the toroidal field is confined in the internal region defined by the last closed poloidal field line. Model B is a crustal confined magnetic field, with an extended (in the angular direction), quadrupolar toroidal field. The upper panels of Fig.~\ref{fig_initial} show the global initial MF geometry, which in both cases matches with an external vacuum dipolar solution, with $B_p^0=10^{14}$~G. The toroidal field is $B_t^0=10^{14}$~G at its maximum. As we are interested in the evolution in the crustal region, for visualisation purposes we show in the lower panels (and hereafter) a planar representation of the MF in the crust.

When the temperature of the NS drops below the melting temperature \citep{baiko96} a solid crust is formed. Within few days, the inner crust up to the neutron drip density is crystallized, while the outer crust is formed within weeks or months \citep{vanriper91,aguilera08,ho12}. The evolution of the MF in the solid crust is described by the Hall induction equation that, including relativistic corrections, has the following form
\begin{equation}
\frac{\partial \vec{B}}{\partial t} = - \vec{\nabla}\times \left[  \eta \curlB - { \vec{v}_e } \times (e^\nu \vec{B} )  \right]~,
\label{induction}
\end{equation} 
where $\eta$ is the magnetic diffusivity, and $\vec{v}_e$ is the {``electron fluid"} velocity given by
 \begin{equation}
 \vec{v}_e = - \frac{e^{-\nu} c }{4\pi e n_e} \curlB
 \label{vel}
\end{equation}
with $n_e$ being the electron number density. The differential operator $\nabla$ associated to the spatial metric includes the corresponding metric factors, i.e. , $e^{-\lambda(r)} \frac{\partial}{\partial r}$. In the core, the field is practically frozen because of the high conductivity of nuclear matter. In the solid crust, in the absence of accretion, the evolution of magnetic field is governed by Ohmic dissipation (the first term on the right hand side of Eq.~(\ref{induction})) and the Hall term (term proportional to $\vec{v}_e$).

To compare the relative importance of the two terms, we denote by $\omega_B\tau_{e}=cB/4\pi en_e \eta$ the
so-called ``magnetization parameter'', usually found in the literature of plasma physics, where $\omega_B$ is the gyro-frequency and $\tau_{e}$ the electron relaxation time. Note that this quantity gives an estimate of the ratio between the characteristic Ohmic dissipation timescale $\tau_{d}=L^2/\eta$ and Hall timescale $\tau_{h}= 4\pi e n_e L^2/c B = \tau_{d}/(\omega_B \tau_e)$. This ratio strongly depends on temperature. Typical values in neutron stars vary in a large range $\tau_d\sim 10^4-10^9$ yr and $\tau_h \sim (10^4-10^6) \frac{10^{14} G}{B}$ yr. In previous works \citep{pons07,pons09,vigano12}, we have discussed in detail the timescales, the mathematical character 
Eq. (\ref{induction}) and the numerical method to solve it.

Following \citep{geppert99}, we introduce a new velocity term ($\vec{v}_e  \rightarrow \vec{v}_e+\vec{v}_a$) in the induction equation to model the effect of accretion. This new term represents the sinking velocity of the accreted matter as it piles up on top of the crust, with
\begin{equation}\label{v_accr}
 \vec{v}_a=-\frac{\dot{m}(\theta)}{\rho}e^{-\nu}\hat{r}
 \label{vela}
\end{equation}
where $\dot{m}$ is the mass accretion rate per unit normal area, which in general depends on the latitude ($\theta$), and $\hat{r}$ is the unit vector in the radial direction. This expression is a result of the continuity equation (conservation of mass) together with the assumption that the accreted mass steadily piles up pushing the crust toward the interior without spreading (pure radial displacements, thus with $\dot{m}$ constant at any depth within a vertical column). 

%%%%%%%%%%%%%%%%
\section{Magneto-rotational evolution}

Our code follows the evolution of the MF inside the star. We impose as a boundary condition at the surface the continuous matching with general vacuum solutions \citep{vigano12}. From the values of the MF at the surface, we can reconstruct the external configuration. In particular, we can follow the evolution of the external dipolar component, $B_p$, responsible for the spin-down of the star. 

%%%%%%%%%%%%%%% 
\begin{figure}
 \centering
\includegraphics[width=.35\textwidth]{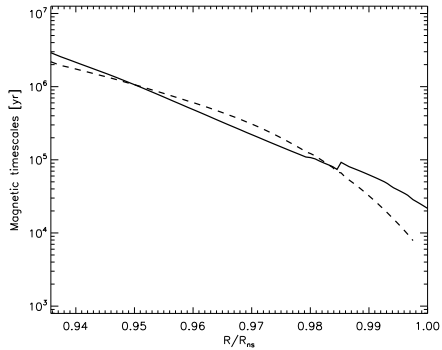}
\caption{Radial profile at the equator of the Ohmic timescale (solid) and Hall timescale (dashed) for initial model A, with $B_p^0=10^{14}$~G and $L=1$ km.} 
 \label{fig_timescales}
\end{figure}

\begin{figure}
 \centering
\includegraphics[width=.4\textwidth]{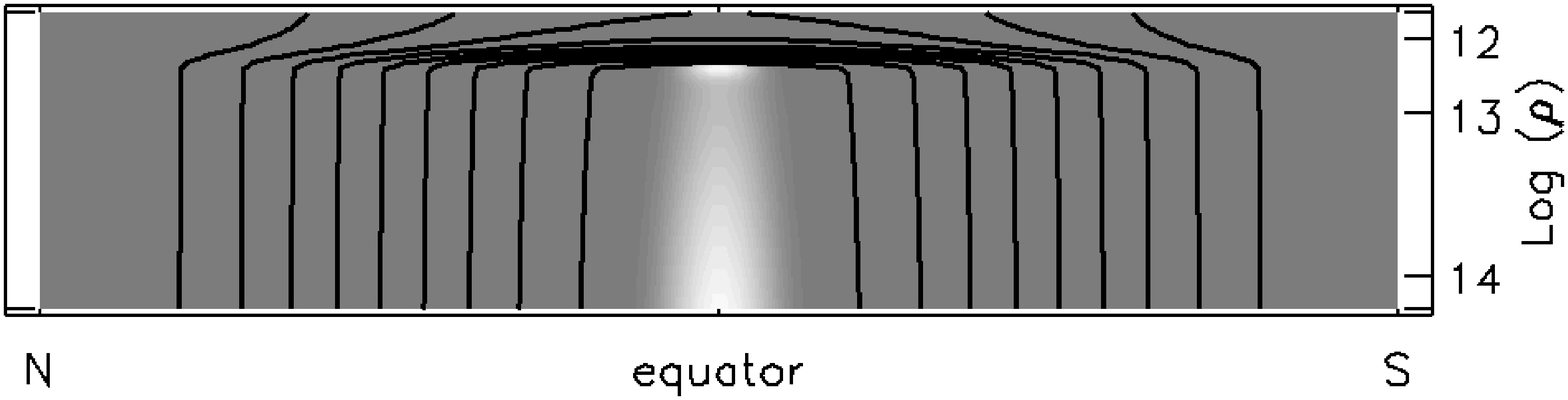}
\includegraphics[width=.4\textwidth]{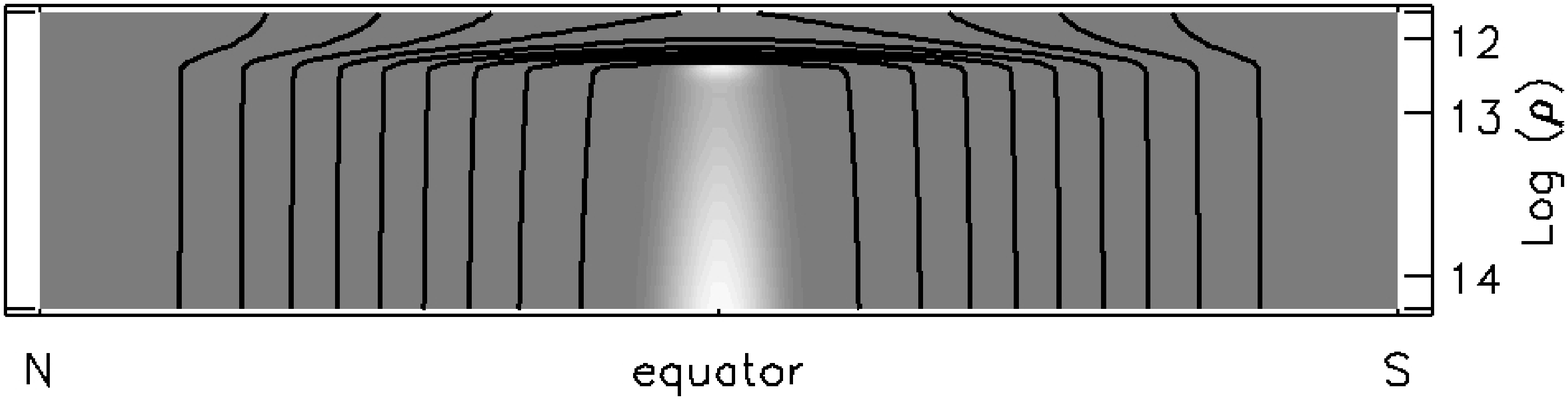}
\includegraphics[width=.4\textwidth]{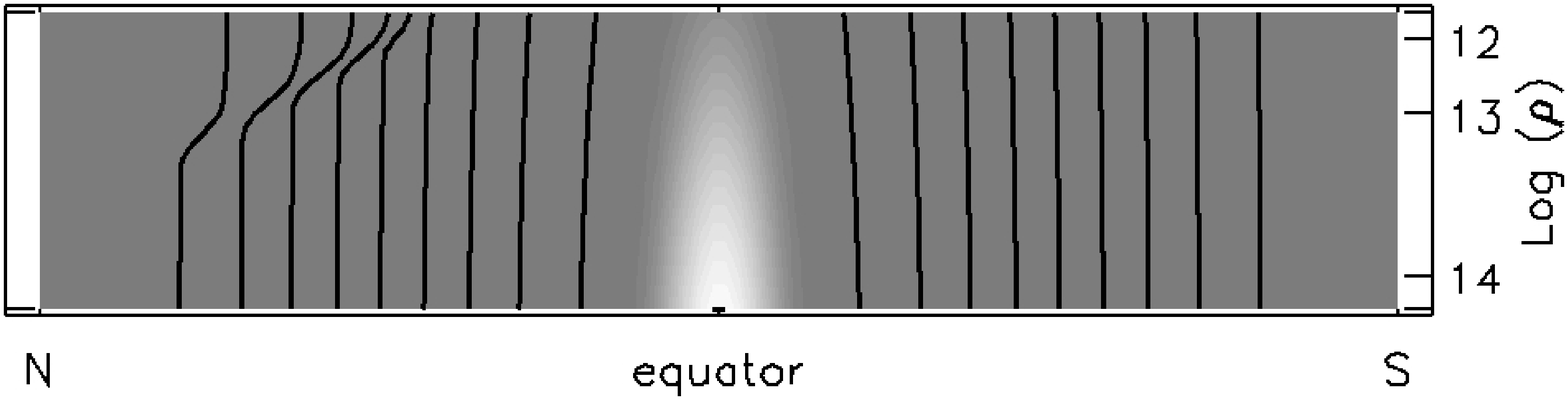}
\includegraphics[width=.4\textwidth]{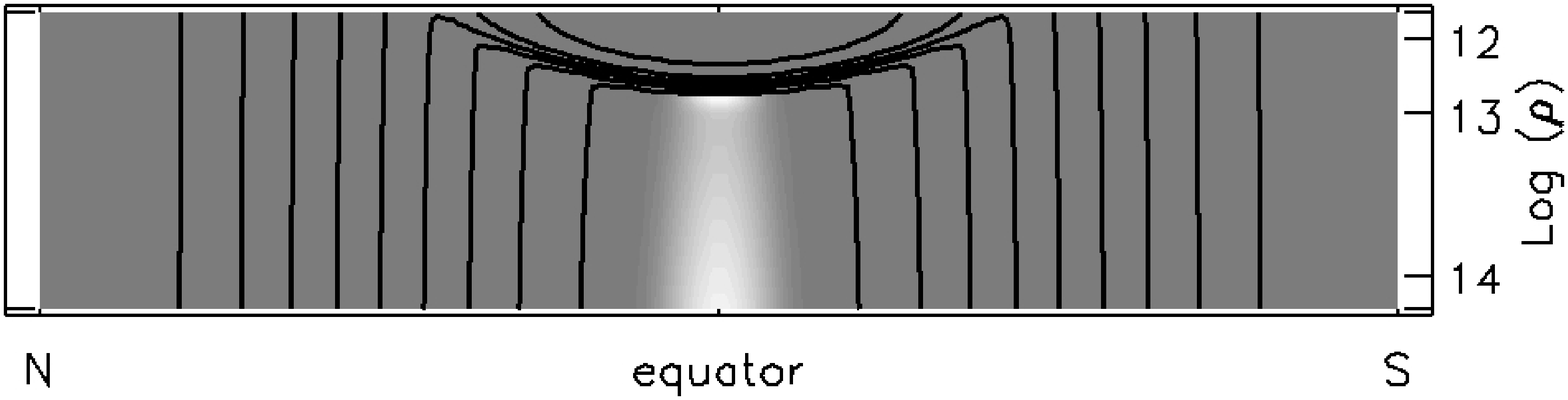}
\includegraphics[width=.4\textwidth]{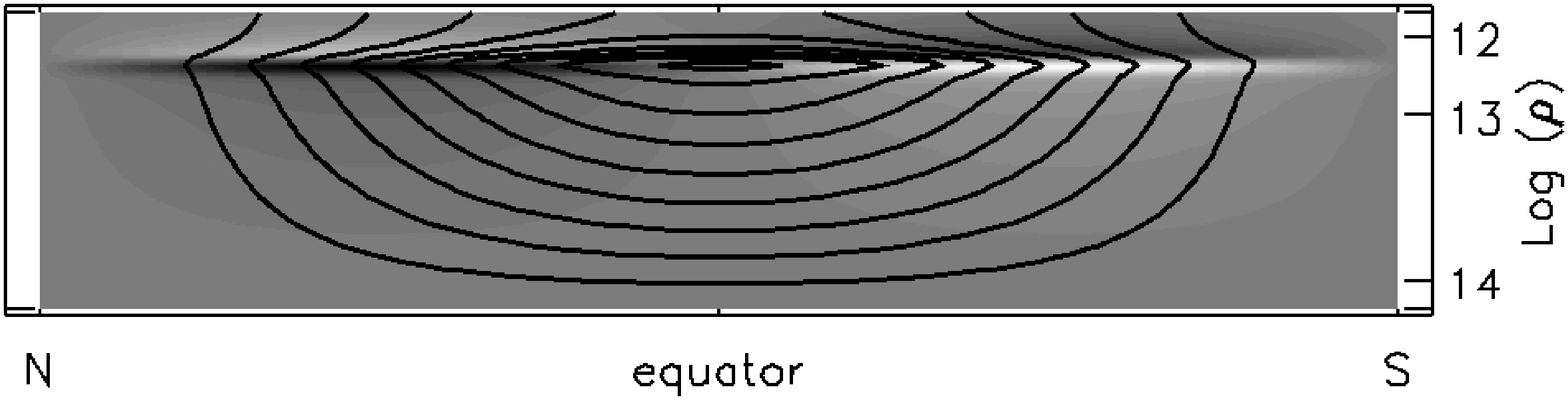}
\caption{Magnetic field in the solid part of the crust after accreting $10^{-4} M_\odot$. Upper panel: model A with $B_p^0=10^{12}$~G and spherical accretion. Second, third, fourth panels: model A with $B_p^0=10^{14}$~G, and (s), (p) and (e) accretion rate geometry, respectively. Fifth panel: model B with $B_p^0=10^{14}$~G and 
spherical accretion.} 
 \label{fig_bury}
\end{figure}

%%%%%%%%%%%%%%
\subsection{Submergence of the magnetic field.}

We study the accretion phase by exploring the sensitivity of the results to two key parameters: the total accreted mass, $M_a$, and the angular dependence of $\dot{m}(\theta)$ in Eq.~(\ref{v_accr}). Neither the final geometry nor the submergence depth at the end of the accretion stage depend on the duration and time dependence of the fallback accretion period that follows a supernova explosion, because of its short duration ($\sim$ months), much shorter than the other relevant timescales shown in Fig.~\ref{fig_timescales}.

It should be noted that our simulations follow the submergence of the MF in the solid crust, as matter piles up on top of the crust, not the dynamical process of accretion outside. This would require MHD simulations of the accretion process in the exterior (see e.g. \cite{bernal10}), which is out of the scope of this paper. Thus, we assume that any details about the dynamics of the accretion process is included in the phenomenological form of $\dot{m}(\theta)$, that describes the rate at which matter reaches the top of the crust at each latitude. How the complex external dynamics leads to a particular form of $\dot{m}(\theta)$ should be the purpose of separate studies. We also note that at the relatively high temperatures ($2-3 \times 10^9$ K) during the early accretion stage in our simulations, fast thermonuclear burning is likely to bring the composition close to the ground state, unlike what happens in accreting, old, cold NSs in binary systems. Therefore we assume that our crust is composed by matter in the ground state.

We begin by discussing how our results depend on the value of $M_a$ for three representative cases:\\
($s$) spherically symmetric, $\dot{m}(\theta)=k$;\\
($p$) channeled onto a polar cap, $\dot{m}(\theta)=ke^{-(\theta/\theta_w)^2}$;\\
($e$) channeled onto the equator, $\dot{m}(\theta)=ke^{-(\frac{\theta-\pi/2}{\theta_w})^2}$;\\
where $k$ is the normalization that fixes $M_a$. Hereafter we discuss our results with $\theta_w=0.4$ rad. In Fig.~\ref{fig_bury} we compare the MF configurations in the crustal region immediately after accretion stops, for models with $M_a=10^{-4} M_\odot$. We look first at the simple case of spherical accretion (upper two panels) for $B_p^0=10^{12}$~G and $B_p^0=10^{14}$~G for model A. During the accretion stage, field lines are advected toward the interior and screening currents are developed in the outer layers. The reduction in the surface field strength is compensated by the local amplification of $B_\theta$ at the submergence depth, where it can reach values up to few $10^{16}$~G. The geometry at the end of the submergence phase is not affected by the initial field strength (it simply scales with $B_p^0$), since the evolution is governed by the term in $\vec{v}_a$ (Eq.~(\ref{vela})), in Eq.~(\ref{induction}). However, anisotropic accretion flows cause the irregular submergence of the MF (middle and bottom panels). Strong screening currents also appear, but now they are localized in the regions where $\dot{m}(\theta)$ is larger. This has an important consequence: the external, large-scale, dipolar component  is not reduced as much as in the spherically symmetric case.

In Table~\ref{tab_sub_reem} we summarize our results for model A, for different values of $M_a$ and varying the accretion geometry, as listed above. Here $\rho_d$ denotes the density at the submergence depth and $B_p^s$ the final strength of the dipolar component of the MF (in all models $B_p^0=10^{14} G$). We point out that the exact value of $B_p^s$ is very sensitive to the boundary conditions, i.e. the magnetic flux exchange with the exterior. We match with a general vacuum solution, while a more consistent approach should consider again the MHD accretion process in the exterior. On the contrary, we found that our results for $\rho_d$, $B_p^r$ and reemergence timescales are not sensitive to the external boundary conditions. For spherical accretion ($s$), we found that if $M_a \lesssim 10^{-5} M_\odot$, the submergence depth is shallow. We observe a sharp transition from $M_a<10^{-4} M_\odot$, with at most a factor 3 decrease in the surface strength, to $M_a>10^{-3} M_\odot$, characterized by a very deep submergence, well within the inner crust.

%%%%%%%%%%%%%%%%%%%%%%%%%%%%%%%%%
\begin{table}
\begin{center}
\footnotesize
 \begin{tabular}[ht!]{c c c c c c}
 geom. & $M_a$ 		& $\rho_d$	& $B_p^s$      & $B_p^r$  \\
       & [$M_\odot$]		& [g cm$^{-3}$] & [10$^{14}$G] & [10$^{14}$G]  \\
\hline 
 s & $2\times 10^{-3}$ & $ 3.0\times 10^{13}$ & $\sim 0$  & 0.76  \\
 s & $1\times 10^{-3}$ & $ 1.6\times 10^{13}$ & $10^{-5}$ & 0.84  \\
 s & $5\times 10^{-4}$ & $ 9.0\times 10^{12}$ & 0.005     & 0.90  \\
 s & $2\times 10^{-4}$ & $ 4.5\times 10^{12}$ & 0.12      & 0.93  \\
 s & $1\times 10^{-4}$ & $ 2.5\times 10^{12}$ & 0.34      & 0.95  \\
 e & $1\times 10^{-4}$ & $ 6.0\times 10^{12}$ & 0.52      & 0.92  \\
 p & $1\times 10^{-4}$ & $ 3.5\times 10^{13}$ & 0.96      & 0.98  \\
 s & $1\times 10^{-5}$ & $ 1.5\times 10^{11}$ & 0.86   	 & 0.98

 \end{tabular}
\caption{Properties for different accretion geometries and total accreted mass ($M_a$) for model A, with $B_p^0=10^{14}$~G. The density at which the maximum compression of lines is reached is $\rho_d$. $B_p^s$ and $B_p^r$ stand for the values of $B_p$ at the end of the submergence stage and the maximum value reached during reemergence, respectively. }
\label{tab_sub_reem}
\end{center}
\end{table} 

%%%%%%%%%%%%%%
Model B (fifth panel) shows a similar behaviour: the submergence depth is the same as model A. No differences in the submergence process arise from the geometry or intensity of the MF: the details of the $M_a-\rho_d$ relation depend mainly on the particular structure and mass of the NS. As a matter of fact, in agreement with the induction and continuity equations, the enclosed mass between $\rho_d$ and the surface is $M_a$, so that for $M_a\gtrsim 10^{-2} M_\odot$ the accreted mass has completely replaced the outer crust.

The rotational behaviour of the star would also be radically different, because the external dipole (third column) regulates the spindown of the star. For $M_a \gtrsim 10^{-3} M_\odot$, and spherical (or nearly isotropic) accumulation of matter on top of the crust, $B_p$ is largely reduced and therefore the spindown rate of the star becomes unusually small. Although estimates of the MF based on timing properties suggest a very low field, the internal field could be orders of magnitude larger,  since it is simply screened by currents in the crust that will be dissipated on longer timescales. Note also that equatorial accretion ($e$) produces a similar reduction of the external dipolar field as for spherical accretion, but the submergence is rather anisotropic (deeper in the equator). On the contrary, accretion concentrated in the polar cap ($p$) barely modifies the MF strength or geometry.

We have also found that adding an azimuthal component of velocity $v_{\phi}\neq 0$ in Eq.~(\ref{v_accr}) does not affect the submergence process, driven solely by the radial inflow. Instead, its effect is the twisting of field lines and the creation of toroidal field, which in turn enhances the Hall activity on longer timescales, during the reemergence process.

\subsection{Reemergence of the magnetic field.}

%%%%%%%%%%%%%%%
\begin{figure}
 \centering
\includegraphics[width=.4\textwidth]{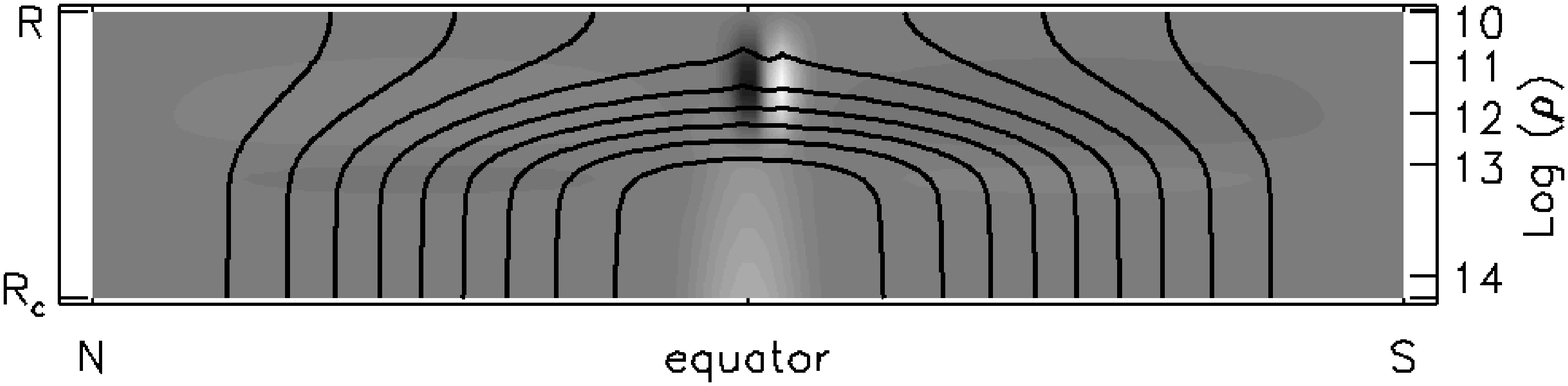}
\includegraphics[width=.4\textwidth]{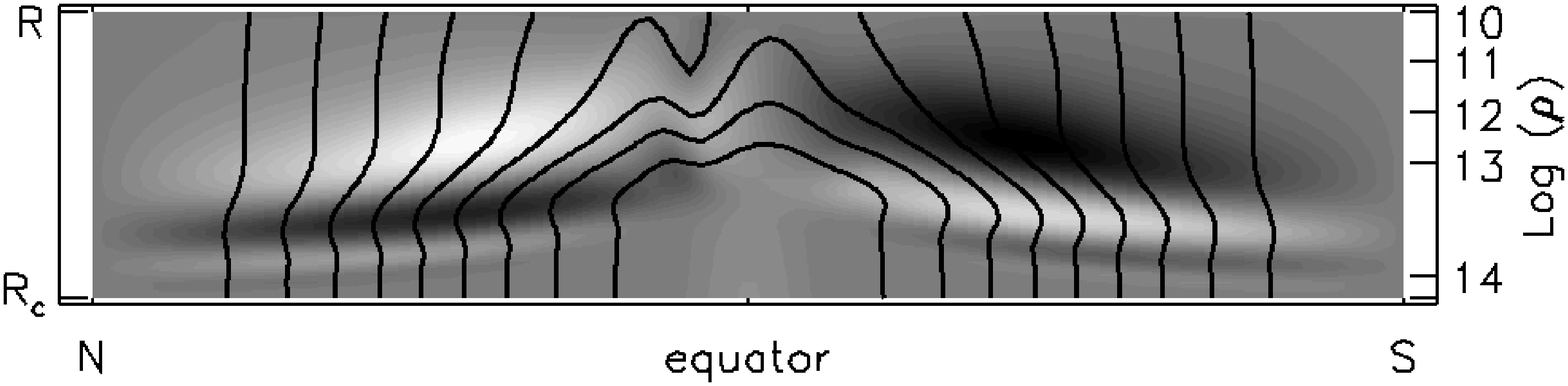}
\includegraphics[width=.4\textwidth]{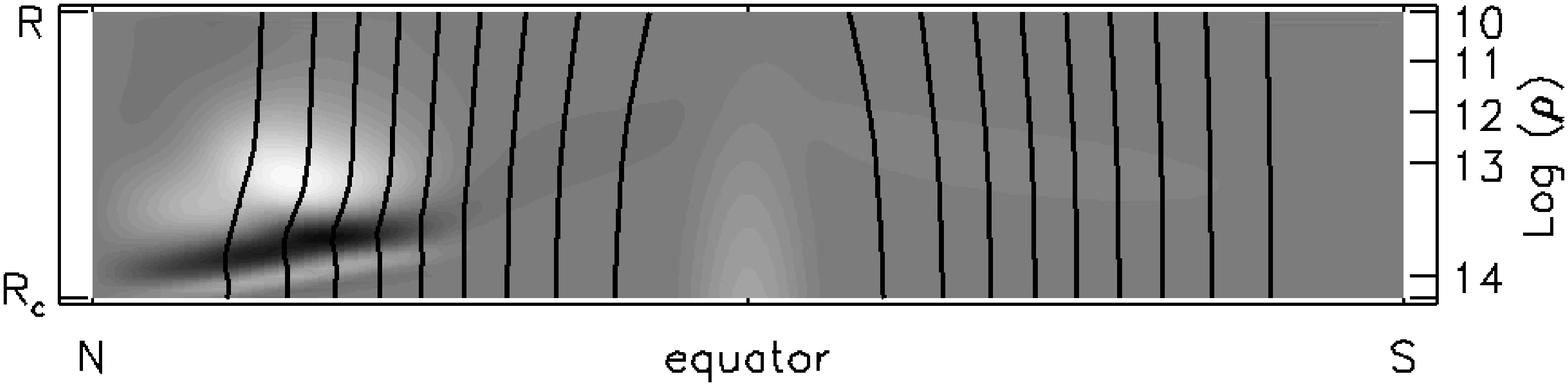}
\includegraphics[width=.4\textwidth]{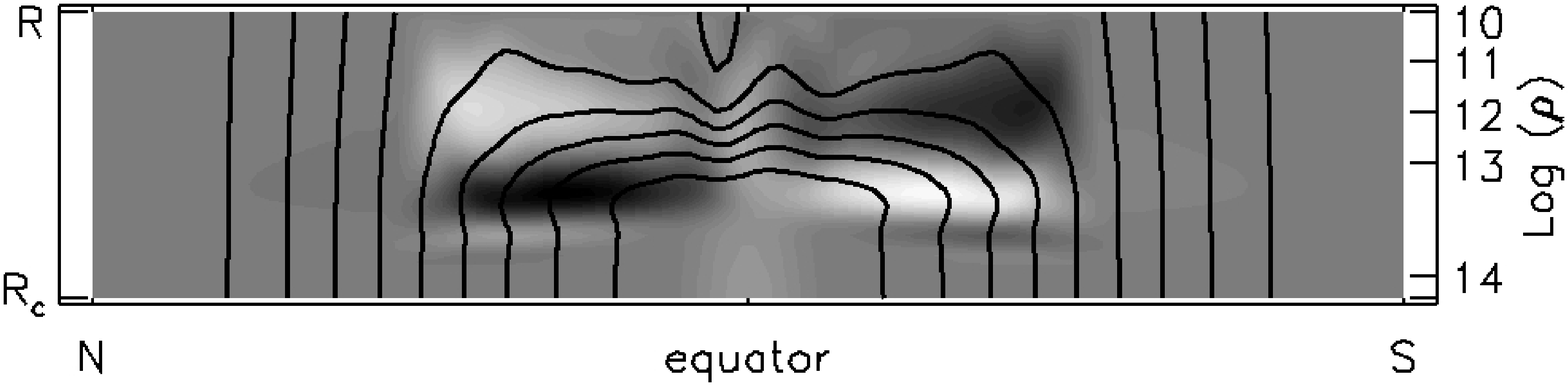}
\includegraphics[width=.4\textwidth]{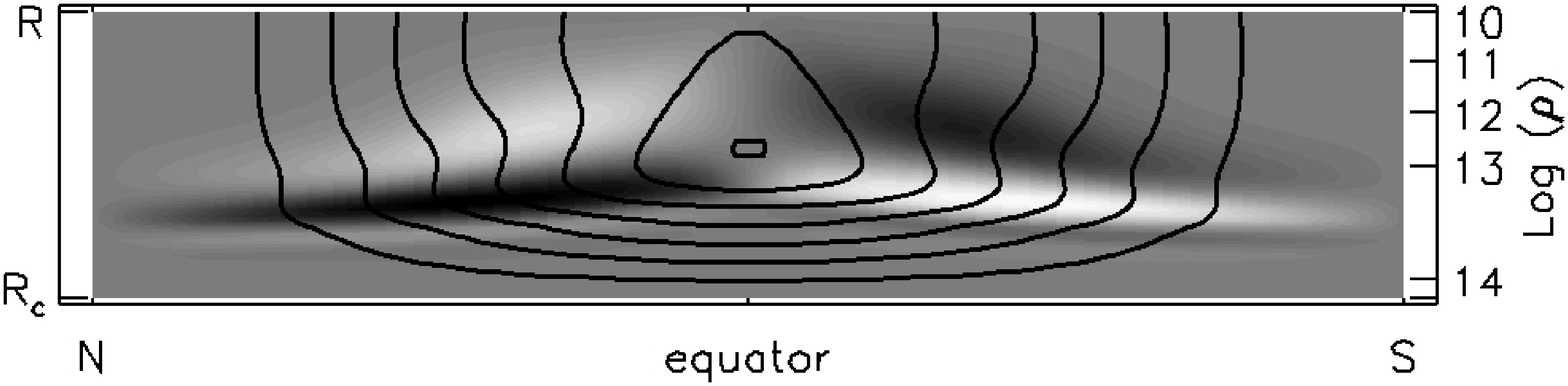}
\caption{Reemergence of MF. Configurations at $t=5$ kyr after the accretion episode, for the same cases shown in Fig.~\ref{fig_bury}.} 
 \label{fig_reemergence}
\end{figure}

After accretion stops, the reemergence process begins. It is mainly driven by Ohmic diffusion (screening currents are concentrated in the outer crust, where the resistivity is relatively high), but also by the Hall term when the MF strength locally exceeds $2-3 \times 10^{14}$~G.
In this latter case, there may be significant changes in the geometry of the MF, including the generation of higher order multipoles in both poloidal and toroidal components, and the launching of whistler waves towards the poles \citep{pons07,vigano12}.

In Fig.~\ref{fig_reemergence} we show the MF configuration at $t=5$ kyr for the same five models as in Fig.~\ref{fig_bury}. In the weak field case (top panel) the field simply reemerges by diffusion. It shows very little Hall activity, seen as generation of toroidal field (white/black spots) and deformation of poloidal field in a small equatorial region of the outer crust. In the other four cases ($B_p^0=10^{14}$~G), the local generation of strong toroidal field and the creation of non-trivial structures are clearly seen, but these small scale features are mainly localized in the inner crust, which makes difficult to predict possible observational consequences. At the same time, the global, large-scale field at the surface is being restored to a shape closer to the original geometry shown in Fig.~\ref{fig_initial}, with additional small scale multipoles near the surface.

\begin{figure}
 \centering
\includegraphics[height=.19\textwidth]{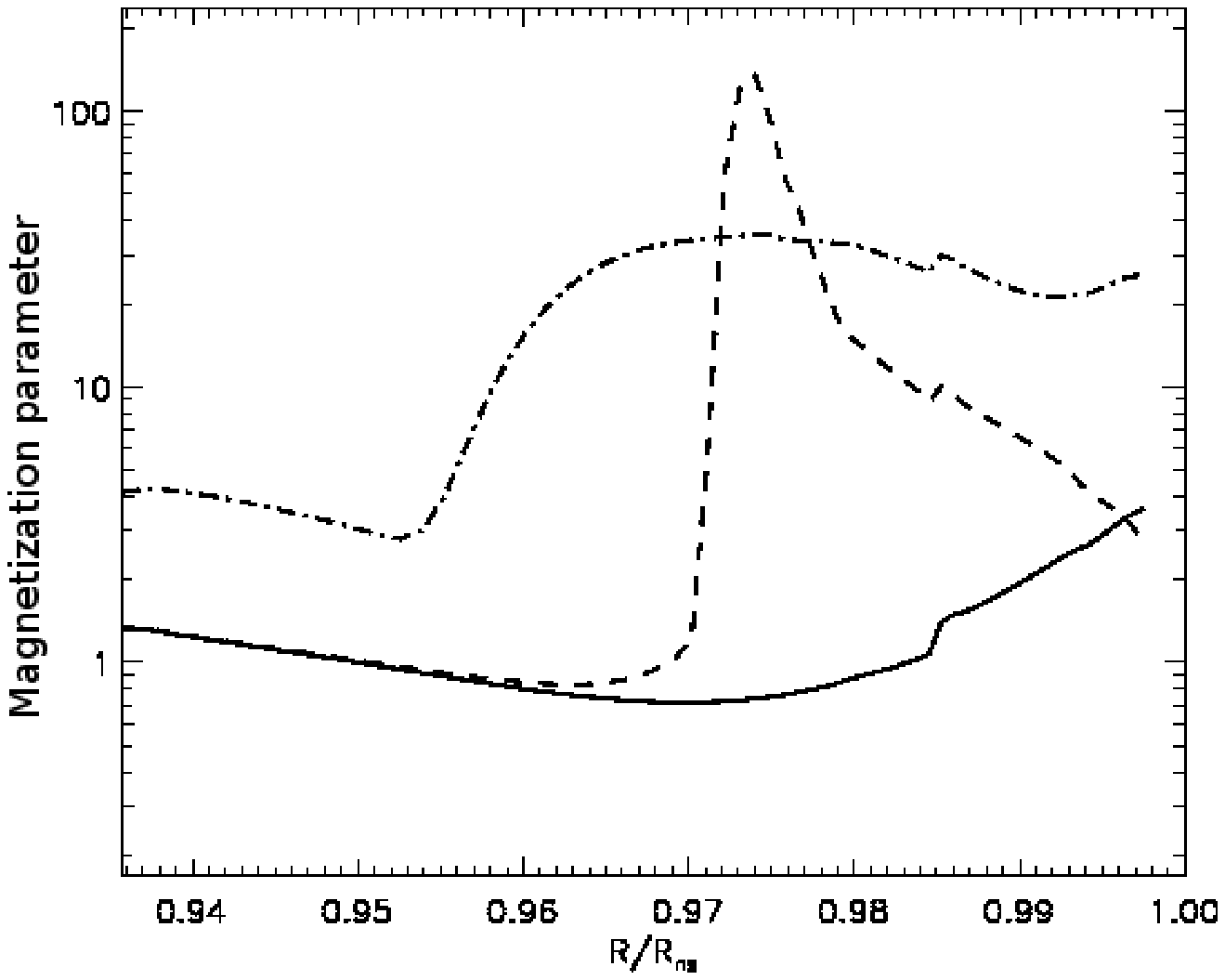}
\includegraphics[height=.19\textwidth]{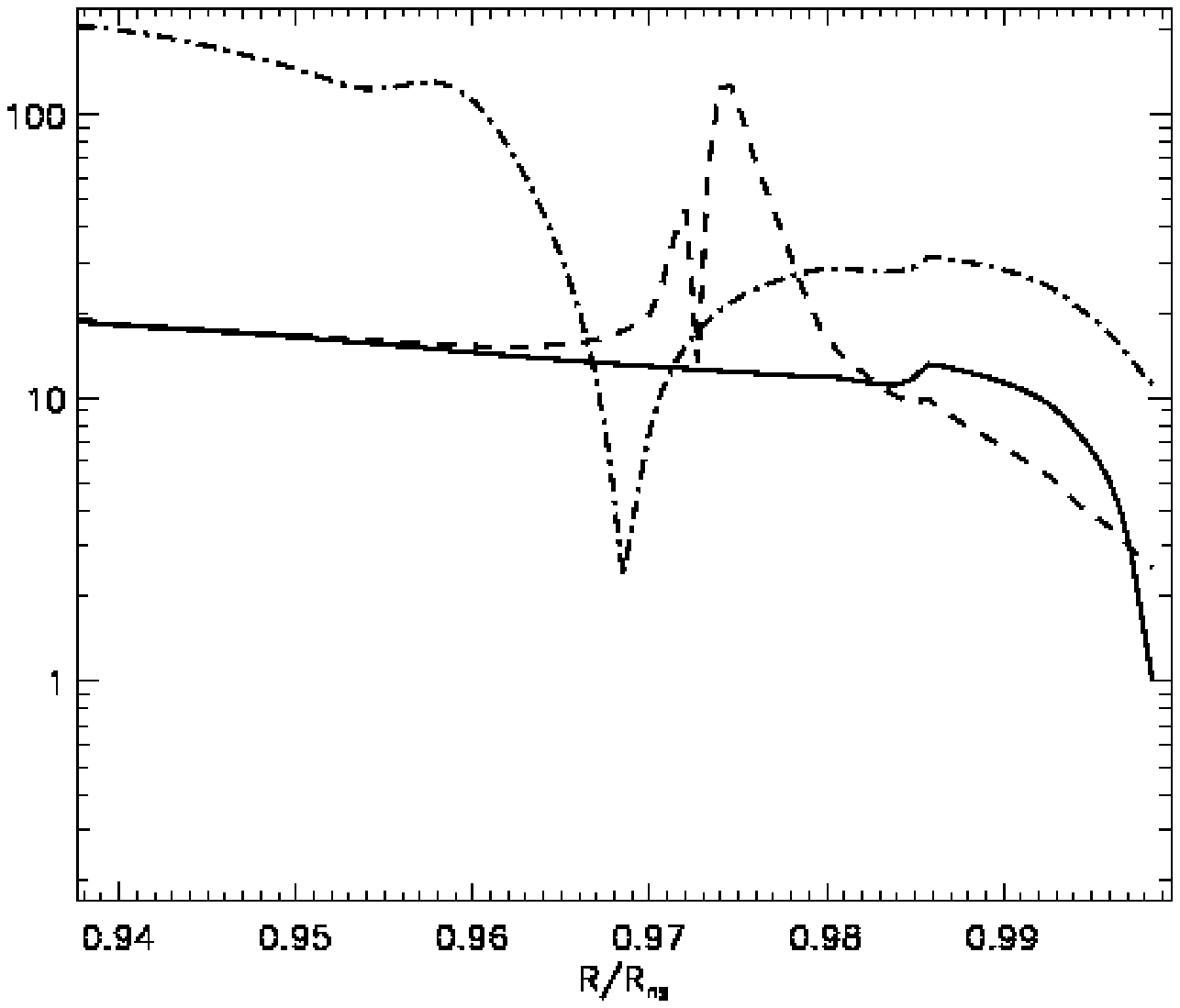}
\caption{Radial profile at the equator of $\omega_B \tau_e$ for initial model A (left) and model B (right), with $B_p^0=10^{14}$~G, at three times: at the beginning (solid), after spherical accretion of $10^{-4} M_\odot$ (dashed) and at $t=5$ kyr, during reemergence (dot-dash).} 
 \label{fig_reynolds}
\end{figure}

The importance of Hall activity is shown in Fig.~\ref{fig_reynolds}. Here we plot the radial profile of the magnetization parameter at the equator, at different stages for model A (left panel) or model B (right panel), both with $B_p^0=10^{14}$~G. The solid line corresponds to the initial configuration, for which $\omega_B\tau_e$ is initially of order 1 for model A, while it is ten times larger for model B, due to the larger mean value of MF. After the submergence (dashed), in both models $\omega_B\tau_e$ reaches much larger values in the region where the field has been compressed. After 5 kyr, the diffusion of the field and the lower temperature make $\omega_B\tau_e>1$ in the whole crust. As $\omega_B\tau_e$ scales with $B$, the same models, but with $B_p^0=10^{12}$~G, give a rescaling of $\omega_B\tau_e$ by two orders of magnitude, providing $\omega_B\tau_e\lesssim 1$ always. 
Thus, for such initially low fields almost no Hall activity is expected.

%%%%%%%%%%%%%%%

\begin{figure}
 \centering
\includegraphics[width=.35\textwidth]{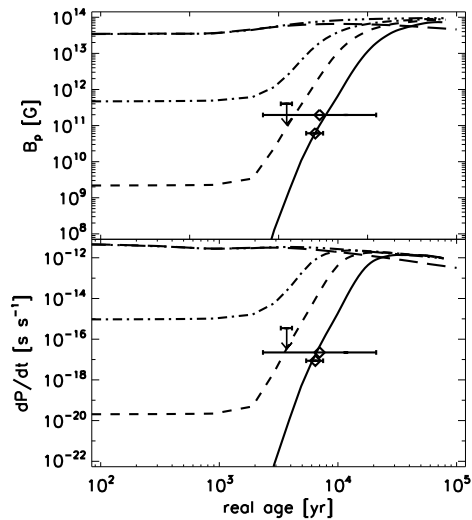}
\caption{Evolution of $B_p$ and $\dot{P}$ during the reemergence phase after spherical accretion, for $B_p^0=10^{14}$~G, assuming $P_0=0.1$ s. Different cases are shown: model B, $M_a=10^{-4}M_\odot$ (long dashes); model A, $M_a=[1,5,10,20]\times 10^{-4} M_\odot$ (other lines from top to bottom, respectively). Observational data are shown with two romboids and an arrow (upper limit). Estimates and errors of the ages of the associated supernova remnants are shown \citep[and refs. within]{ho11}. } 
 \label{fig_timing}
\end{figure}

In the top panel of Fig.~\ref{fig_timing} we show the evolution of $B_p$ as a function of time after a spherical fallback episode, with different $M_a$, for model B (long dashes) or model A (other linestyles), together with the inferred values (or upper limit) of observed CCOs. The {\it reemergence timescale}, on which the surface MF grows, depends basically on $\rho_d$. For total accreted matter in the range of interest, there is an initial delay of about $10^3$ yr before appreciable reemergence is observed. The reemergence process takes between $10^3$ and few $10^5$, and it is determined by the local conditions (in particular, the resistivity) where the screening currents are located. The MF will never reach the original strength, since some Joule dissipation is always expected. Generally speaking, $B_p$ is restored to close to its initial value for the core-extended configurations (model A). Only for the extreme case of $M_a \gtrsim 10^{-2} M_\odot$, when the reemergence process lasts a long time, $B_p^r$ is significantly reduced (see Table 1). In model B (long dashes), crustal currents are much stronger than in model A, so both Ohmic and Hall timescales are shorter. This implies that the reemerged field has been more  dissipated, in agreement with 1D studies \citep{geppert99,ho11}. Furthermore, as the field is more compressed, the Hall activity is more intense (see dot-dash line in Fig.~\ref{fig_reynolds}), and the transfer of magnetic energy from the dipolar component to higher multipoles (i.e., small scale structures) is faster.

The reemergence stage has a strong imprint on timing properties. Assuming a small period at the end of the accretion stage, $P_0=0.1$ s, we follow the evolution of the timing properties ($P$ and $\dot{P}$) using the classical formula for an orthogonal rotator: $B_p=(6.4 \times 10^{19} \sqrt{ P\dot{P}I_{45}}/R_{10}^3)$~G, where $R_{10}=R/10$ km and $I_{45}=I/10^{45}$ g cm$^2$ are given by the radius and moment of inertia of the star model. The bottom panel of Fig.~\ref{fig_timing} shows the evolution of $\dot{P}$ as a function of the real age for several values of $M_a$ in the sensitivity range $[10^{-4}-10^{-3}] M_\odot$. Variations of a factor of a few in $M_a$ within this range result in an extremely low value of $\dot{P}$ during the first thousands of years of a NS life. In terms of the spin-down age, $\tau_c=P/2\dot{P}$, this means that it overestimates the real age also by the same factor. Observational data reported for CCOs are consistent with $M_a \gtrsim10^{-3} M_\odot$.

\begin{figure}
 \centering
\includegraphics[width=.35\textwidth]{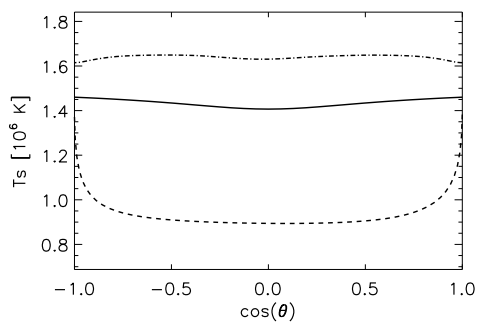}
\caption{Comparison between surface temperature distributions of an anti-magnetar (solid), and two models of hidden magnetars (dashed and dot-dashed) after accreting $M_a=10^{-3} M_\odot$. All three models are shown in an evolutionary phase when the external dipolar field is $B_p=10^{10}$~G.} 
 \label{fig_ts}
\end{figure}

A possible observable to distinguish between an ``anti-magnetar" and the hidden magnetic field scenario is the surface temperature distribution, $T_s(\theta)$, that determines the average luminosity and light-curve in the X-ray band. Fig.~\ref{fig_ts} shows $T_s(\theta)$ for two high field NSs during the reemergence phase after a $10^{-3} M_\odot$ accretion episode and an antimagnetar. The dot-dashed line refers to model A (Fig.~\ref{fig_initial}) at $t_1=3$ kyr, while the dashed line corresponds to a crustal-confined magnetic field model, with a larger toroidal component ($B_p^0=10^{13}$~G, $B_t^0=10^{15}$~G), at $t_2=6$ kyr. In both cases, the field has partially reemerged and has same value as for the  antimagnetar model (solid lines), $B_p=10^{10}$~G. Model A has a slightly warmer surface than the antimagnetar model, but both are almost isotropic. On the contrary, the crustal confined model shows a distinctive feature consisting in a lower average $T_s$ with hot polar caps. This is produced by the buried toroidal field that keeps the equatorial surface region insulated from the warmer core. The anisotropy depends on the accreted mass, the natal magnetic field and it varies with time depending on the location of currents during the reemergence process.

%%%%%%%%%%%%%%%
\section{Final remarks}

We have presented results from detailed simulations of one of the possible scenarios proposed to explain the low inferred external MFs of CCO. For the first time, a consistent, coupled magneto-thermal evolution in 2D and including the Hall term has been used to model the evolution of the magnetic field in both submergence and reemergence  phase, under the effect of different accretion flow geometries. Our study confirms the qualitative results obtained by previous works \citep{geppert99} for 1D simplified models, in what concerns the relation between submergence depth with total accreted mass. However, for large magnetic fields, the dynamics of the reemergence phase is more complex in our models, since we are not restricted to one particular component (dipole) and the interplay between toroidal and poloidal field is allowed.

We have confirmed that the {\it hidden magnetar} model is a possible solution, but it requires that, immediately following the supernova explosion, $10^{-3} M_\odot$ are able to reach a large fraction of the the top of the crust. Under this circumstances, the accretion flow is able to push the field lines deep inside the crust. Other choices, with channeled accretion onto a polar cap or onto the equatorial region, have a more local effect that will barely modify the large-scale MF. If the NS is born as a typical pulsar ($10^{12}$~G), rather than a magnetar, a reduction of only 1-2 orders of magnitude of the external MF is easier accomplished with less accreted matter. \cite{chevalier89} 
suggested a value of $M_a\sim 0.1 M_\odot$ for SN 1987A, while $M_a\lesssim 10^{-3} M_\odot$ is expected for a more typical type II SN. Interestingly, this estimate lies in our range of interest. Independent work modeling the MHD accretion process in the exterior of a NS surrounded by the dirty environment of the supernova explosion is needed to provide more realistic boundary conditions and to explain the required values of $M_a$ for a deep submergence of the field.

We have also explored the subsequent evolution of the crustal MF during reemergence by Ohmic dissipation of screening currents. If the hidden MF is of the order of $10^{14}$~G, non-linear dynamics driven by the Hall term are also expected, with the generation of toroidal field and multipolar components. This activity is, however, localized in the inner crust, and it remains unclear if the surface field (and therefore timing properties) can reflect in some way this non-trivial structures. In any case, the reemergence process by diffusion has a clear imprint on the timing properties of the NS. The temporary low values of $\dot{P}$ may point to weak inferred MFs, but that is simply an artifact of the complex field geometry. The reemergence phase, for the explored range of $M_a$, may last from $10^3$ to $10^5$ years, after which we expect that the field has almost completely reemerged, although it is partially dissipated. Another important observational consequence of this scenario is the effect on the braking index, $n$. During reemergence, the field appears to be growing to an external observer, which means $n<3$. Interestingly, this is in agreement with all reported values measured for young pulsars ($<10^4$ yrs) \citep{hobbs10,espinoza11}, perhaps pointing to hypercritical accretion in the aftermath of supernova explosions as a very common scenario.

Last, the thermal evolution during the reemergence phase in the hidden magnetic field scenario can yield to totally different degrees of anisotropy in the surface temperature, depending on the initial MF and on the accreted mass. A hidden magnetar with a large, extended toroidal field buried in the crust could qualitatively explain the surface temperatures anisotropies inferred in several CCOs: the large pulsed fraction observed in Kes 79 \citep{halpern10,shabaltas12}, the antipodal hot spots seen in Puppis A \citep{deluca12}, and the small emitting region of the blackbody components needed to fit the spectra of 1E 1207 \citep{deluca04}. Instead, weaker (or less extended) hidden toroidal MF provide low pulsed fractions.

We conclude that the hidden magnetic field scenario for CCOs is not in contradiction with any observational data, and it would not be surprising that one of this objects displays magnetar-like activity in the form of bursts or flares, as it has happened for the {\it low-field magnetar} SGR 0418+5729 \citep{rea10,turolla11}. We also consider the possibility that not all the CCO candidates are of the same nature: large, magnetar-like hidden MFs are particulary suitable for those CCOs who are displaying temperature anisotropies. On the other hand, some of the standard, young rotation-powered pulsars (RPPs) could be in their final stage of reemergence, representing the connection between CCOs and RPPs/magnetars population.

\section*{Acknowledgments}
This work was partly supported by the Spanish grant AYA 2010-21097-C03-02 and CompStar, a Research Networking Program of the European Science Foundation. DV is supported by a fellowship from the \textit{Prometeo} program for research groups of excellence of the Generalitat Valenciana (Prometeo/2009/103). We thank J.A.~Miralles and S.~Mereghetti for many helpful comments.

\label{lastpage}
\bibliography{cco_accretion}

\end{document}